\documentclass[a4paper,twocolumn,11pt,unpublished]{quantumarticle}
\pdfoutput=1
\usepackage[utf8]{inputenc}
\usepackage[english]{babel}
\usepackage[T1]{fontenc}
\usepackage[numbers,sort&compress]{natbib}
\usepackage{graphicx}
\usepackage{amsmath}
\usepackage{hyperref}
\usepackage[braket, qm]{qcircuit}
\usepackage{tikz}
\usepackage{float}
\usepackage{algorithmic}
\usepackage[section]{algorithm}

\begin{document}

\title{Modeling Linear Inequality Constraints in Quadratic Binary Optimization for Variational Quantum Eigensolver}
\author{Miguel Paredes Qui\~{n}ones*}
\affiliation{IBM Research, 04007-900 São Paulo, Brazil}
\email{mparedes@br.ibm.com}
\author{Catarina Junqueira}
\affiliation{\textit{Linear Softwares Matemáticos}, 01419-002 São Paulo, Brazil}
\email{catarinajunqueira2@gmail.com}

\maketitle

\section{Abstract}
  This paper introduces the use of tailored variational forms for variational quantum eigensolver that have properties of representing certain constraints on the search domain of a linear constrained quadratic binary optimization problem solution. Four constraints that usually appear in several optimization problems are modeled. The main advantage of the proposed methodology is that the number of parameters on the variational form remain constant and depend on the number of variables that appear on the constraints. Moreover, this variational form always produces feasible solutions for the represented constraints differing from penalization techniques commonly used to translate constrained problems into unconstrained one. The methodology is implemented in a real quantum computer for two known optimization problems: the Facility Location Problem and the Set Packing Problem. The results obtained for this two problems with VQE using 2-Local variational form and a general QAOA implementation are compared, and indicate that less quantum gates and parameters were used, leading to a faster convergence. 

\section{Introduction}
Binary quadratic optimization problems are among the main and well-studied combinatorial problems. The term binary means that the variables can assume only two possible values, which are usually 0 or 1. In ising models \cite{10.3389/fphy.2014.00005}, these variables could be -1 or 1. In this paper the following formulation is used for a Linear Constrained Quadratic Binary Optimization Problem (LCQBO):\\

\begin{align}
\label{eq:binaryop}
    \min_{x} &\; \; x^{T} Q x \\
    s.t. &  Ax = b \\
    &  Bx \leq c\\
      &   x \in \{0,1 \} 
\end{align}

Classically, this problem is solved with global convergence by using Branch and Bound, which is not efficient when the number of variables increase, since the search tree could grow exponentially. As an alternative, some heuristics with lower guaranty are used.

Heuristics based on quantum computing have recently been developed to solve the unconstrained version of the LCQBO, called Quadratic Unconstrained Binary Optimization (QUBO). Some of these heuristics are the Variational Quantum Eigensolver (VQE) \cite{Peruzzo2014,Li2011} and its special form called Quantum Approximation Algorithm (QAOA)\cite{farhi2014quantum}. The VQE was initially applied on quantum chemistry problems, such as the problem of searching for molecular ground states (minimum energy of the system). Once QUBO problems can be mapped as an ising formulation, which is one of the formulations for the Hamiltonian calculation, it is possible to solve the QUBO with VQE \cite{Moll2017} and to approximate the solution of an LCQBO by using penalisation techniques, reducing it to QUBO~\cite{10.3389/fphy.2014.00005,glover2018tutorial}.


Another approach for solving the QUBO is to use quantum computation based on the quantum adiabatic theorem. This approach allows ground states calculations but it has limitations when solving QUBO problems \cite{Smolin2013}. 

The use of variational forms for expressing Ansatze solutions for the VQE algorithm had an important contribution on the development of an efficient VQE solver \cite{Kandala2017}. Despite that, there is still a challenge in using the classical solver part of VQE, which is to ensure the expression of every solution in the search domain that is parameterized by means of rotation angles and cnot gates \cite{Shende2003}.

For quantum chemistry problems there are tailored strategies for representing Ansatze as an Unitary Coupled Cluster (UCC) \cite{Lee_2018} and its Singles and Doubles Excitation Variational Form (UCCSD) \cite{Barkoutsos_2018}. For these strategies the Ansatz only contains single and double ex-citation operators, which potentially decrease the number of parameters for representing states. Some of these Ansatze are particularly developed to work efficiently with hardware \cite{Kandala2017}, even though they still have the limitation of over-sampling the Hilbert space \cite{McClean_2018}. There is also the dynamic creation of Ansatze, where an operator is added to the Ansatz at every step, from a pre-defined set of operators, this technique is called Adaptive Derivative Assembled Pseudo-Trotter (ADAPT-VQE)~\cite{Grimsley_2019}. In this paper we compare our solution with 2-Local circuits as Ansatze, which is an heuristic circuit to map the Hilbert space that assumes that qubits have almost 2 local qubits, and then they can be circularly entangled to represent the system \cite{Sim_2019}.


Taking this into account, this paper aims to address this difficulty when VQE is formulated for LCQBO with the construction of special variational forms in a way that the search domain attain specific constraints. In \cite{matsuo2020problemspecific} is presented a methodology to prepare different types of constraints from those proposed in this paper.

This paper is organized as follows, In Section~\ref{sec:Methods} we presents a review of the VQE in order to understand the position of our contribution on the development of an efficient VQE solver for LCQBO problems, and circuit cost criteria to compare other Ansatze. Then in Section~\ref{sec:Taylored} the development of the variational forms in order to attain certain constraints for the search space domain is explained. Next, Section~\ref{sec:application} shows the use of the variational forms in two different optimization problems: the Facility Location Problem (FLC) and the Linear Assignment Problem (LAP). Finally, in Section \ref{sec:Conclusions} we discuss the findings  and some future work are outlined.  

\section{Methods}\label{sec:Methods}
\subsection{Variational Quantum Eigensolver}
\label{sec:VQE}
The VQE is a hybrid algorithm that has a quantum part and a classical part \cite{McClean_2016}. It is a type of near-term algorithm that uses noisy quantum computers to calculate expectation values of a minimum energy state. Originally, the VQE was used in quantum chemistry to approximate the minimum estate of energy of a quantum system represented by a Hamiltonian using the Variational theorem. Then, some class of  classical optimization problems called QUBO problem could be mapped on like Hamiltonians. Next, Algorithm~\ref{al:VQE} presents the overall procedure for a generic VQE for solving QUBO problems.

\begin{algorithm}[H]
    \caption{General Variational Quantum Eigensolver}
   \label{al:VQE}
    \begin{algorithmic}[1]
    \REQUIRE Set Number of shots $N_{shots}$
    \REQUIRE Give initial state $\ket{\psi_{0}}$
    \REQUIRE Map quadratic objective function $Q$ to Pauli gates $H_{i}$. 
    \WHILE{classical optimization condition}
    \STATE Construct Ansatz $U(\theta_{k})$
    \STATE Apply Ansatz to the initial state:\\
    $ \ket{\psi(\theta_{k})}= U(\theta_{k})\ket{\psi_{0}}$
    \STATE Measure output $\ket{\psi(\theta_{k})}$ $N_{shots}$ times \\
    \STATE Calculate the expected value of the objective function $ E \left( \sum_{i} \bra{\psi(\theta_{k})} H _{i}\ket{\psi(\theta_{k})} \right)$
    \STATE Classical optimization algorithm update $\theta_{k}$.
    \ENDWHILE
    \STATE Optimal set of parameters $\theta^{*}$
\end{algorithmic}
\end{algorithm}

In Algorithm~\ref{al:VQE} we observe that first it is needed to define the number of times that the Ansatz circuit will be executed at one iteration of the VQE. Then, it is necessary to inform an initial state $\ket{\psi_{0}}$. In quantum chemistry, this state usually is calculated with a classical procedure like the Hartree–Fock method~\cite{FROESEFISCHER1987355}. To solve optimization problems there is no clear strategy to calculate this initial state, but as in quantum chemistry, this could be done by getting a solution of a fast classical heuristic or by a relaxation method that leads to a good quality feasible solution. 

Since we are trying to minimize a classical QUBO problem we need to map the objective function to diagonal Hamiltonians $H_{i}$ represented by Pauli gates. First we use ising representations of QUBO problems \cite{10.3389/fphy.2014.00005}. One practical way to make this transformation is to replace every $x_{i}$ variable on the original classical problem by $(1-Z_{i})/2$, where $Z_{i}$ is a pauli gate applied in qubit $i$ with $+-1$ eigenvalues on its matrix representation. Since we can translate quadratic objective functions to diagonalized Hamiltonian it is possible to use penalization strategy to convert a linear equality constraint as $Ax = b$ into $\lambda \|Ax - b \|_{F}$ on the objective function \cite{10.3389/fphy.2014.00005}.

The following step is to determine the Ansatz that we will be used to sample the Hilbert space of solutions of our problem. Usually to do it, a parameterized Ansatz $U(\theta)$ with wide potential to sample Hilbert space is created, while $\theta$ angles varies. One desirable characteristic of the Ansatz is to use less single and double qubit gates. This preference is because current quantum hardware has a limitation on the depth of the circuits that can be executed, and also a limited number of entanglement between qubits.

Then, the Ansatz is sampled a number of times for a fixed set parameters $\theta_{k}$. With this we can calculate an expected value of the energy $E\left[ \sum_{i} \bra{\psi(\theta_{k})} H_{i} \ket{\psi(\theta_{k})} \right]$. This expected value of the energy for state $\ket{\theta_{k}}$ represent an approximation for the objective function of the original classical optimization problem. This is based on the Variational method in quantum mechanics where the minimum energy and others energy states have a relation \eqref{eq:variational}, and is equal only when the state $\ket{\psi_{g}}$ correspond to the ground state energy $E_{g}$.

\begin{equation} \label{eq:variational}
    E\left[ \sum_{i} \bra{\psi(\theta_{k})} H_{i} \ket{\psi(\theta_{k})} \right] \geq E_{g}
\end{equation}

The final component of the VQE algorithm is the classical optimization algorithm, where the parametric energy is minimized $\bra{\psi(\theta)} H \ket{\psi_{\theta}}$ with a classical solver by updating parameters $\theta$. In \cite{Sung_2020}, many of the following solvers are explored and its effectiveness on considering or not noise on the objective function. Some of the most used solvers are: Simultaneous   Perturbation Stochastic Approximation (SPSA) \cite{119632}, the Nelder-Mead Simplex Method \cite{10.1093/comjnl/7.4.308}, Sequential Least Squares Programming (SLSQP) \cite{dai2008} and Adam \cite{adam2014}. One improved algorithm for objective functions with trigonometric functions is Nakanishi-Fujii-Todo algorithm (NFT)~\cite{Nakanishi_2020} solver, which is also robust against statistical error. One limitation of the parametric circuits in a classical optimization loop is that it creates extremely non-linear optimization problems where there exists plateaus around local minimum and also from where it is difficult to escape \cite{McClean_2018}. For our computational experiments we use Constrained Optimization by Linear Approximations (COBYLA)\cite{Powell1994}.

\subsection{Circuits costs}
All the variational forms proposed in this paper uses $R_{y}$ gates with parameters $\theta$. State of the art approaches for create an Ansatz variational form uses $R_{y}$ and $R_{z}$ rotations conjugated with cnot gates. Then it is repeated the Ansatz an empirical number of times , this is named the "depth" of the Ansatz \cite{Moll2017}. When VQE is implemented and defined a depth, its created a set with polynomial number of parameters  \cite{Peruzzo2014}. This set of parameters will be the search domain when is passed to the classical optimization problem  part \cite{Sim_2019}. We need to use a metric to compare the cost of execute the Ansatze that we use on section~\ref{sec:application}.\\

\begin{table}[h]
    \centering
    \begin{tabular}{|l|c|c|c|}
\hline
  Var. Form   &  \# $SU(2)$  & \# cnots & \# param.\\
  \hline
    Fig~\ref{fig:circuit_selection} & $2N-1$ & $N-1$ & $N$  \\
    Fig~\ref{fig:circuit_disjunctive} &$2N-1$ & $2(N-1)$ &  $N$ \\
    Fig~\ref{fig:circuit_combination} &$2N-1$ & $4N-6$ &  $N$ \\
    Fig~\ref{fig:circuit_combination2} &$2N-3$ & $3N-5$ &  $N-1$ \\
\hline
    \end{tabular}
    \caption{Number of cnots, $SU(2)$ gates and parameters to execute TVFs.}
    \label{tab:costs}
\end{table}
On table~\ref{tab:costs} is shown the number of single-gates and cnots to run tailored variational forms. This decomposition on basic gates is relevant since with the current Noisy Intermediate-Scale Quantum (NISQ) devices, noises impact more on the use of cnot gates, then a efficient quantum algorithm should avoid to use this gates when possible. It is possible to approximate it with a relation of that cnot gates takes 10 times more than a single-qubit gate as is showed in  \cite{zhang2018efficient}.

\begin{equation}
    cost=N_{CNOT} \times 10 +N_{SU(2)}
\end{equation}

\section{Taylored Variational Forms (TVF)}
\label{sec:Taylored}
In this section we will create parameterized circuits for variational forms for representing specific constraints. For the starting point we will choose $\ket{\psi}_{0} = \ket{0}^{\otimes N}$, where $N$ is the number of qubits, and apply a quantum circuit represented by the gate $U_{c}$ for obtaining only feasible solutions for our binary problems.

\subsection{Binary comparison constrains}
\label{sec:conditional}
The main idea is to model the behavior of a binary comparison sequence of constraints, such as:
\begin{equation} \label{eq:conditional}
    x_{i} \leq x_{i+1} \; \forall i = 0 \ldots N 
\end{equation}
with $x_{0} = 0$. The possible solutions for the variables $x_{i}$ and $x_{i+1}$ are shown in Table\ref{tab:possible}.

\begin{table}[h]
    \centering
    \begin{tabular}{|c|c|c|}
\hline
  $x_{i}$   &  $x_{i+1}$  & Possible \\
  \hline
    0 & 0 & yes \\
    0 & 1 & yes \\
    1 & 0 & no \\
    1 & 1 & yes \\
\hline
    \end{tabular}
    \caption{Possible solutions for constraint~\eqref{eq:conditional}}
    \label{tab:possible}
\end{table}

Therefore, it is possible to code the variables $x_{i}$ on $N$ qubits states as in equation \eqref{eq:representation}.
\begin{equation} \label{eq:representation}
    \ket{x_{1} \; \ldots \; x_{i} \; \ldots \; x_{N} }
\end{equation}{}

If all qubits states from Table \ref{tab:possible} were generalized, then only the states with the following sequence would be obtained.

\begin{equation}
      \ket{ 0 \ldots 00 } , \ket{ 0 \ldots 0 1 } , \ket{ 0 \ldots 1 1 } ,\ldots  , \ket{ 1 \ldots 11 }  \label{eq:desired2}
\end{equation}

As an alternative, it is possible to use the following non-conventional qubit representation with integers on ket notation, as shown in equation \eqref{eq:desired}.
\begin{equation}
      \ket{0} , \ket{ 2^{N-1}} , \ket{ 2^{N-1}+2^{N-2}} ,\ldots  , \ket{ \sum_{j=N-i}^{N-1}2^{j} }  \label{eq:desired}
\end{equation}

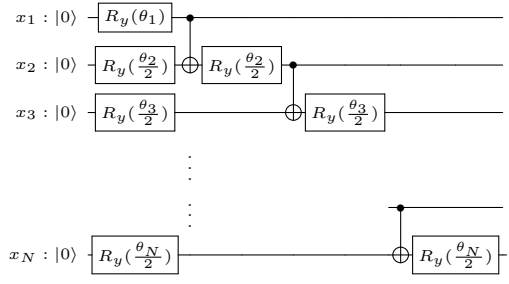
\begin{figure}[h]
    \centering
    \tiny
    \begin{equation*}
    \Qcircuit @C=0.2em @R=0.5em @!R {
	 	\lstick{ x_{1} : \ket{0} } & \gate{R_y(\theta_{1})} & \ctrl{1} & \qw& \qw  & \qw & \qw & \qw & \qw & \qw \\
	 	\lstick{ x_{2} : \ket{0} } & \gate{R_y(\frac{\theta_{2}}{2})} & \targ & \gate{R_y(\frac{\theta_{2}}{2})} & \ctrl{1} & \qw & \qw& \qw & \qw& \qw \\
	 	\lstick{ x_{3} : \ket{0} } & \gate{R_y(\frac{\theta_{3}}{2})} & \qw & \qw & \targ & \gate{R_y(\frac{\theta_{3}}{2})}  & \qw & \qw & \qw & \qw \\
	 		& & \vdots & &   \\
	 	& & \vdots & & & & &\ctrl{1} & \qw & \qw  \\
	 	\lstick{ x_{N} : \ket{0} } & \gate{R_y(\frac{\theta_{N}}{2})} & \qw & \qw & \qw & \qw & \qw & \targ &\gate{R_y(\frac{\theta_{N}}{2})} & \qw & \qw
	 }
\end{equation*}
    \caption{Variational form with $\theta_{i}$ parameters to represent all possible $x_{i}$ states.}
    \label{fig:circuit_selection}
\end{figure}

If the variational form in Figure~\ref{fig:circuit_selection} were applied to a \ket{0}, the result would be the desired superposition of states shown in \eqref{eq:desired}, with the amplitude probabilities parameterized on angles $\theta_{i}$.  


{}


Where $R_{y}$ is a single-qubit rotation through $\theta$ angle in radians around the y-axis.


To illustrate how this circuit only gets answers as in \eqref{eq:desired}, with probabilities depending on $\theta_{i}$ angles, let us apply the circuit for the first 2 qubits and observe the sequence formed. The states initialize in $\ket{0}^{\otimes N }$, then $R_{y}(\theta_{i})$ is applied for every qubit and the following is obtained.
\begin{align}
    &\left( \cos{\frac{\theta_{1}}{2}}\ket{0} +  \sin{\frac{\theta_{1}}{2}} \ket{1} \right) 
    \otimes \notag \\
    &\left( \cos{\frac{\theta_{2}}{4}} \ket{0} +  \sin{\frac{\theta_{2}}{4}} \ket{1} \right)  \label{eq:1ev } \\
    \rightarrow & \left( \cos{\frac{\theta_{1}}{2}}\cos{\frac{\theta_{2}}{4}} \ket{00} + \cos{\frac{\theta_{1}}{2}} \sin{\frac{\theta_{2}}{4}} \ket{01} \right. \notag\\
    & \left. + \sin{\frac{\theta_{1}}{2}}\cos{\frac{\theta_{2}}{4}} \ket{10} +\sin{\frac{\theta_{1}}{2}}\sin{\frac{\theta_{2}}{4}} \ket{11} \right)\otimes \ldots
\end{align}{}

Now, applying the cnot gate with qubit 1 as a control and qubit 2 as a target:
\begin{align}
    &\left( \cos{\frac{\theta_{1}}{2}}\ket{0} +  \sin{\frac{\theta_{1}}{2}} \ket{1} \right) 
    \otimes \notag \\
    &\left( \cos{\frac{\theta_{2}}{4}} \ket{0} +  \sin{\frac{\theta_{2}}{4}} \ket{1} \right)  \label{eq:3ev } \\
    \rightarrow & \left( \cos{\frac{\theta_{1}}{2}}\cos{\frac{\theta_{2}}{4}} \ket{00} + \cos{\frac{\theta_{1}}{2}} \sin{\frac{\theta_{2}}{4}} \ket{01} \right. \notag\\
    & \left. + \sin{\frac{\theta_{1}}{2}}\cos{\frac{\theta_{2}}{4}} \ket{11} +\sin{\frac{\theta_{1}}{2}}\sin{\frac{\theta_{2}}{4}} \ket{10} \right)\otimes \ldots
\end{align}{}
Then, let us apply $R_{y}(\theta/2)$ gate to the second qubit:

\begin{align}
    \rightarrow \left( \cos{\frac{\theta_{1}}{2}}\cos{\frac{\theta_{2}}{2}} \ket{00} + \cos{\frac{\theta_{1}}{2}} \sin{\frac{\theta_{2}}{2}} \ket{01} \right. \notag \\
    + \left. \sin{\frac{\theta_{1}}{2}} \ket{11} \right)\otimes \ldots \label{eq:2ev}
\end{align}{}

By induction from equations \eqref{eq:1ev } and \eqref{eq:2ev}, one can observe that if the operation is performed for $N$ qubits then equation \eqref{eq:validstates1} is obtained.

\begin{equation} \label{eq:validstates1}
    \ket{\psi(\mathbf{\theta})} = \sum_{i=0}^{N} \left( \prod_{k=i+1}^{N} \cos{\frac{\theta_{k}}{2}}\right) \sin{\frac{\theta_{N-i+1}}{2}} \ket{\sum_{j=N-i}^{N-1}2^{j}}
\end{equation}{}

with $\theta_{N+1}=\pi$. As can be seen in \eqref{eq:validstates1}, $N$ parameters $\theta$, $N$ gates $R_{y}$ and $N-1$ controlled $R_{y}$ gates are needed to represent $N+1$ states in $\ket{\psi}$. \\

\subsection{Binary sum less than one constrains}  
\label{sec:disjunctive}
Similarly as in Section~\ref{sec:conditional} we will deduce a variational form for represent constraints, in this case we represent only one constraint~\eqref{eq:disjunctive} for a set of variables, then:

\begin{equation} \label{eq:disjunctive}
    \sum_{i=1}^{N} x_{i} \leq 1
\end{equation}
We examine the possible solutions that the variables $x_{i}$ and $x_{i+1}$ can attain in Table\ref{tab:possible}.

\begin{table}[h]
    \centering
    \begin{tabular}{|c|c|c|}
\hline
  $x_{i}$   &  $x_{i+1}$  & Possible \\
  \hline
    0 & 0 & yes \\
    0 & 1 & yes \\
    1 & 0 & yes \\
    1 & 1 & no \\
\hline
    \end{tabular}
    \caption{Possible answers complaining equation~\eqref{eq:disjunctive}}
    \label{tab:possible2}
\end{table}
We use the same representation as in~\eqref{eq:representation} Then in a similar way if we generalise states from Table\ref{tab:possible2} we want to obtain only states with the following sequence:

\begin{equation}
      \ket{ 0 0\ldots 0 } , \ket{ 10 \ldots 0 } , \ket{ 01 \ldots 0 } ,\ldots  , \ket{ 0 \ldots 1 }  \label{eq:desired2}
\end{equation}

or using the following non conventional qubit representation:
\begin{equation}
      \ket{0} , \ket{ 2^{0}} , \ket{ 2^{1}} ,\ldots  , \ket{ 2^{i-1} }  \label{eq:desired3}
\end{equation}

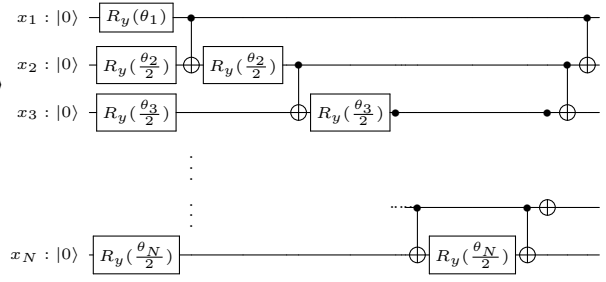
\begin{figure}[h]
    \centering
    \;
    \tiny
    \begin{equation*}
    \Qcircuit @C=0.2em @R=0.5em @!R {
	 	\lstick{ x_{1} : \ket{0} } & \gate{R_y(\theta_{1})} & \ctrl{1} & \qw & \qw & \qw & \qw & \qw & \qw & \qw & \qw & \qw & \qw & \qw & \qw & \ctrl{1} & \qw \\
	 	\lstick{ x_{2} : \ket{0} } & \gate{R_y(\frac{\theta_{2}}{2})} & \targ &\gate{R_y(\frac{\theta_{2}}{2})} & \qw &  \ctrl{1} & \qw& \qw & \qw & \qw & \qw & \qw & \qw & \qw & \ctrl{1} & \targ  & \qw \\
	 	\lstick{ x_{3} : \ket{0} } & \gate{R_y(\frac{\theta_{3}}{2})} & \qw &  \qw &\qw  & \targ & \gate{R_y(\frac{\theta_{3}}{2})} &\ctrl{0} & \qw & \qw & \qw & \qw &  \qw & \ctrl{0} & \targ & \qw & \qw\\
	 		& & \vdots &   \\
	 	& & \vdots & & & &  & &  \cdots & \cdots & \ctrl{1}& \qw & \ctrl{1} & \targ & \qw & \qw & \qw \\
	 	\lstick{ x_{N} : \ket{0} } & \gate{R_y(\frac{\theta_{N}}{2})}& \qw  & \qw & \qw & \qw & \qw & \qw &\qw &\qw & \targ &  \gate{R_y(\frac{\theta_{N}}{2})} & \targ & \qw & \qw & \qw & \qw
	 }
\end{equation*}
    \caption{Variational form with $\theta_{i}$ parameters to represent all possible  $x_{i}$ states for equation \eqref{eq:disjunctive} }
    \label{fig:circuit_disjunctive}
\end{figure}

If we use the following variational form in Fig \ref{fig:circuit_disjunctive} give us the desired states \eqref{eq:desired2} with amplitude probabilities depending on angles $\theta_{i}$.


{}
We apply gates to the $\ket{0}$ initial state and perform similar analisis as in \eqref{eq:1ev } and \eqref{eq:2ev} and obtain:

\begin{align} \label{eq:validstates2}
    \ket{\psi(\mathbf{\theta})} = \sum_{i=1}^{N+1} & \left( \prod_{k=1}^{i-1} \sin{\frac{\theta_{k}+ \mathbf{1}_{\{ k=1\}} \pi }{2}}\right) \\
    & \cdot \cos{\frac{\mathbf{1}_{\{ i=1\}}\pi - \theta_{i}}{2}} \ket{\mathbf{1}_{\{N>i\}}2^{i-1}} \notag
\end{align}{}

where: $\theta_{N+1} = 0 $ and $\mathbf{1}_{\{N>i\}}$ is an indicator function such that $\mathbf{1}_{\{N>i\}} = 1 $ if $N>i$ and $0$ otherwise.

\subsection{Other kind of constraints}
Since mapping the possible solutions of a constraint and express it in a circuit, it give us the idea that other constraints can be constructed as is shown in \cite{matsuo2020problemspecific}, where is implemented circuits for the following equality constraints:
\begin{equation}
    \sum_{i=1}^{N} x_{i} = 1
\end{equation}
\begin{equation}
    \left(1-x\right)\left(1-y\right) = 0
\end{equation}

Another possible constraint to be represented is \eqref{eq:combination} by mixing circuits from Figure~\ref{fig:circuit_selection} and Figure~\ref{fig:circuit_disjunctive}.
\begin{equation} \label{eq:combination}
    \sum_{i=1}^{N-1}x_{i} \leq x_{N}
\end{equation}

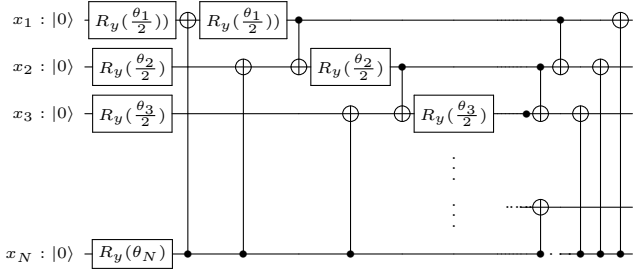
\begin{figure}[h]
    \centering
    \tiny
    \begin{equation*}
    \Qcircuit @C=0.2em @R=0.5em @!R {
	 	\lstick{ x_{1} : \ket{0} } & \gate{R_y(\frac{\theta_{1}}{2}))} & \targ & \gate{R_y(\frac{\theta_{1}}{2}))} & \ctrl{1} & \qw & \qw & \qw & \qw & \qw & \qw & \qw & \qw & \qw & \qw & \qw & \qw & \ctrl{1}  & \qw & \qw & \targ & \qw  \\
	 	\lstick{ x_{2} : \ket{0} } & \gate{R_y(\frac{\theta_{2}}{2})} & \qw & \targ & \targ &\gate{R_y(\frac{\theta_{2}}{2})} &  \ctrl{1} &\qw &  \qw& \qw & \qw & \qw & \qw & \qw & \qw & \qw & \ctrl{1} & \targ   & \qw & \targ & \qw & \qw   \\
	 	\lstick{ x_{3} : \ket{0} } & \gate{R_y(\frac{\theta_{3}}{2})} & \qw &  \qw &  \qw &\targ  & \targ & \gate{R_y(\frac{\theta_{3}}{2})}  & \qw & \qw & \qw & \qw & \qw & \qw &  \qw & \ctrl{0} & \targ & \qw  & \targ & \qw & \qw & \qw \\
	 		& & & & & & & \vdots &   \\
	 	& & & & & & & \vdots & & & &  & &  \cdots & \cdots & \qw & \targ & \qw &\qw & \qw & \qw  & \qw  \\
	 	\lstick{ x_{N} : \ket{0} } & \gate{R_y(\theta_{N})}& \ctrl{-5} & \ctrl{-4} & \qw & \ctrl{-3} & \qw & \qw & \qw &\qw &\qw & \qw & \qw & \qw  & \qw & \qw & \ctrl{-1} & \cdots &  \ctrl{-3} & \ctrl{-4} & \ctrl{-5} & \qw 
	 }
\end{equation*}
    \caption{Variational form with $\theta_{i}$ parameters to represent all possible  $x_{i}, y$ states for equation \eqref{eq:combination} }
    \label{fig:circuit_combination}
\end{figure}{}

Also the constrained version of \eqref{eq:combination}:
\begin{equation} \label{eq:combination2}
    \sum_{i=1}^{N-1}x_{i} = x_{N}
\end{equation}

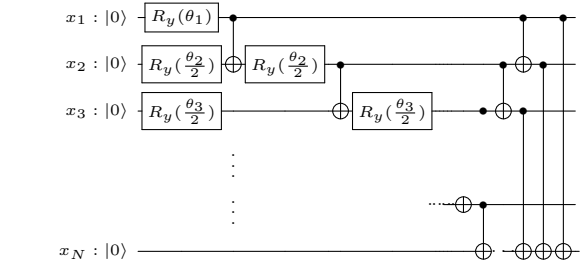
\begin{figure}[h]
    \centering
    \tiny
    \begin{equation*}
    \Qcircuit @C=0.2em @R=0.5em @!R {
	 	\lstick{ x_{1} : \ket{0} } & \gate{R_y(\theta_{1})} & \ctrl{1} & \qw & \qw & \qw & \qw & \qw & \qw & \qw & \qw & \qw & \qw & \qw & \qw & \ctrl{1} &\qw & \ctrl{5} &\qw \\
	 	\lstick{ x_{2} : \ket{0} } & \gate{R_y(\frac{\theta_{2}}{2})} & \targ &\gate{R_y(\frac{\theta_{2}}{2})} & \qw &  \ctrl{1} & \qw& \qw & \qw & \qw & \qw & \qw & \qw & \qw & \ctrl{1} & \targ  & \ctrl{4} & \qw & \qw \\
	 	\lstick{ x_{3} : \ket{0} } & \gate{R_y(\frac{\theta_{3}}{2})} & \qw &  \qw &\qw  & \targ & \gate{R_y(\frac{\theta_{3}}{2})} & \qw & \qw & \qw & \qw & \qw &  \qw & \ctrl{0} & \targ & \ctrl{3} & \qw & \qw & \qw  \\
	 		& & \vdots &   \\
	 	& & \vdots & & & &  & &  \cdots & \cdots & \qw & \qw & \targ & \ctrl{1} & \qw & \qw & \qw & \qw & \qw  \\
	 	\lstick{ x_{N} : \ket{0} } & \qw & \qw  & \qw & \qw & \qw & \qw & \qw &\qw &\qw &  \qw & \qw & \qw & \targ & \cdots & \targ & \targ & \targ & \qw & \qw 
	 }
\end{equation*}
     \caption{Variational form with $\theta_{i}$ parameters to represent all possible  $x_{i}, y$ states for equation \eqref{eq:combination2} }
    \label{fig:circuit_combination2}
\end{figure}{}

All the four circuits developed in this paper are possible to include as constraints for a LCQBO problem. One limitation of this methodology that is not always possible to combine the circuits to represent multiple constrains at the same time, this is a topic for further investigation.

\section{Applications}
\label{sec:application}
All numerical experiments was implemented using IBM quantum computing suite named Qiskit 0.19.2 \cite{Qiskit} with Python 3.7.5. For the classical optimization algorithm we choose COBYLA \cite{Powell1994}, This solver performed better than other state of the art solvers for our test cases. Every calculation of the expected value of the objective function takes 1024 circuit executions for the two methodologies tested. For the initial set of parameters we use random numbers on the interval $[-\pi,\pi]$.

For sake of comparison we use two algorithms one based on VQE using 2-Local variational form and QAOA. Our criteria to set parameters on 2-Local and QAOA are based on the similarity with the number of parameters and single-gates and double-gates on its correspondent TVF. For 2-Local we choose a depth of 1 with $R_{y}$ and cnots as rotation and entanglement blocks. For QAOA we choose $p = 2$ in order to get 4 parameters in total, which are similar to the number of parameters using TVF. As we see, one important decision in the process of implement 2-Local and QAOA is the necessity to estimate meta-parameters, that normally there is no clear criteria to do that.

For model constraints of our test problems using 2-Local and QAOA we use the penalization techniques indicated in \cite{glover2018tutorial} as in shown in table~\ref{tab:penalties}, where $\lambda$ is a penalization factor. 
\begin{table}[h]
    \centering
    \begin{tabular}{|l|l|}
\hline
  Classic Constraint   &  Equivalent Penalty\\
  \hline
     $x \leq y$ & $\lambda\left(x-xy \right)$ \\
    $\sum_{i}^{n} x_{i} \leq 1$ & $\lambda \left( \sum_{i}^{n} \sum_{j < i} x_{i}x_{j}\right)$ \\
\hline
    \end{tabular}
    \caption{Penalization techniques to model some inequality constraints \cite{glover2018tutorial}. }
    \label{tab:penalties}
\end{table}

\subsection{Facility Location Problem}

The Facility Location Problem (FLP) consists of deciding which facilities should be opened from a set of $n$ possible facilities. In various situations, the facilities location decision and the assignment of clients to the facilities are made simultaneously \cite{ARENALES}. This problem can be applied to a numerous types of real problems, such as deciding the location of distribution centers, airports, schools, hospitals, police stations, and others.

To formulate the FLP, let us define $f_{i}$ as the fixed cost of opening facility $i$, for $i=1,...,n$, and $c_{ij}$ as the cost of serving customer $j$ by the facility $i$, for $j = 1,...,m$ and $i=1,...,n$. 

The variables of the FLP are the following:

$y_{i}= \left\{
\begin{array}{ll}
 1 &  \text{\parbox{9.0cm}{if facility $i$ is opened,}} \\
 0 & \text{otherwise.}   \\
\end{array}
\right.$
\newline

$x_{ij}= \left\{
\begin{array}{ll}
 1 &  \text{\parbox{9.0cm}{if facility $i$ serves customer j,}} \\
 0 & \text{otherwise.}   \\
\end{array}
\right.$
\newline

Then, the FLP can be formulated as presented in equations \eqref{eq_objFLP} to \eqref{eqFLP3}.

\begin{align}
    \min_{x,y} & & \sum_{i=1}^{n} f_{i}y_{i}+ \sum_{i=1}^{n}\sum_{j=1}^{m}c_{ij}x_{ij} \label{eq_objFLP} \\
    s.t. & &\sum_{i=1}^{n} x_{ij} = 1 & \; ~~ j = 1,...,m \label{eqFLP2} \\
    & & x_{ij} \leq y_{i} &\; ~~ i = 1,...,n \label{eqFLP3} \\
    & & & \; ~~ j = 1,...,m  \notag 
\end{align}

The objective function \eqref{eq_objFLP} minimizes the total cost of assigning customers to facilities. Constraints \eqref{eqFLP2} guarantee that each customer $j$ is served by exactly one facility, and constraints \eqref{eqFLP3} guarantee that each customer $j$ can only be assigned to a facility $i$ if that facility $i$ is open. For a more detailed review on the FLP and solution techniques, see \cite{FARAHANI2010}.

It is possible to model constraint \eqref{eqFLP2} by adding the penalization $\lambda \sum_{j=1}^{m} \left( 1- \sum_{i=1}^{n}x_{i,j} \right)^2$. We observe that $y_{i}$ variable appears in various constrains \eqref{eqFLP3} with variables $x_{i,j}$, so we can't use circuit in Figure~\ref{fig:circuit_selection}, since this not possible to put all of this constraint at the same time. As an alternative we use a circuit that invert the order of the qubit inequality and permit that the first qubit can control more than 1 qubit as is depicted in Figure~\ref{fig:circuit_FLP}:

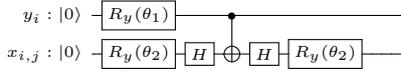
\begin{figure}[h]
    \centering
    \tiny
    \begin{equation*}
    \Qcircuit @C=0.5em @R=0.5em @!R {
	 	\lstick{ y_{i} : \ket{0} } & \gate{R_y(\theta_{1})} & \qw &\ctrl{1} & \qw& \qw  & \qw & \qw & \qw & \qw \\
	 	\lstick{ x_{i,j} : \ket{0} } & \gate{R_y(\theta_{2})} & \gate{H} &\targ & \gate{H} & \gate{R_y(\theta_{2})} & \qw & \qw & \qw& \qw  
	 }
\end{equation*}
    \caption{Reformulated variational form to fit FLP problem characteristics. }
    \label{fig:circuit_FLP}
\end{figure}{}
Then is possible to represent constraints \eqref{eqFLP3} by creating a circuit according to the algorithm~\ref{al:FL}, that generalizes circuit in Figure~\ref{fig:circuit_FLP}.

\begin{algorithm}[H]
    \caption{Facility Location problem Circuit construction}
    \label{al:FL}
    \begin{algorithmic}[1]
    \REQUIRE Number of clients $m$  and facilities $n$.
    \REQUIRE label first $m$ qubits as $x_{i,j}$ $\forall$ $i=1\ldots n $ ; $j =1 \ldots m $
    \REQUIRE label following $n$ qubits as $y_{i}$ $\forall$ $i=1\ldots n $
    \STATE set all qubits with state 0.
   \FOR{$i = 1\ldots n$}
    \STATE apply $R_{y}(\theta_{y_{i}})$ on  $y_{i}$
     \FOR{$j =  1 \ldots m $}
     \STATE apply $R_{y}(\theta_{x_{i,j}})$ on  $x_{i,j}$
     \STATE apply $H$ on  $x_{i,j}$
    \STATE apply $CX$ with $y_{i}$  as a control and $x_{i,j}$  as target
    \STATE apply $H$ on  $x_{i,j}$
   \STATE apply $R_{y}(\theta_{x_{i,j}})$ on  $x_{i,j}$
    \ENDFOR
    \ENDFOR
\end{algorithmic}
\end{algorithm}

As a test we implement the following instance of the facility location problem using our methodology:
\begin{align} 
    \min_{x,y} & &  5y_{1}+10y_{2} + 3x_{1,1} + 2x_{2,1} \label{eq:facility_exampleobj} \\
    s.t. & & x_{1,1}+x_{2,1} = 1 & \label{eq:facility_example1}\\
    & & x_{1,1} \leq y_{1} &  \label{eq:facility_example2}\\
    & & x_{2,1} \leq y_{2} & \label{eq:facility_example3}
\end{align}
This instance includes 2 facilities and 1 client with their respective costs.

We implement this instance of the FL problem with 4 qubits $(q_{1} \leftarrow x_{1,1}, q_{2} \leftarrow x_{2,1}, q_{3} \leftarrow y_{1}, q_{4} \leftarrow y_{2})$. We use a penalty factor of $\lambda = 100$. 

\begin{figure}[h]
    \begin{center}
   \resizebox{\linewidth}{!}{\input{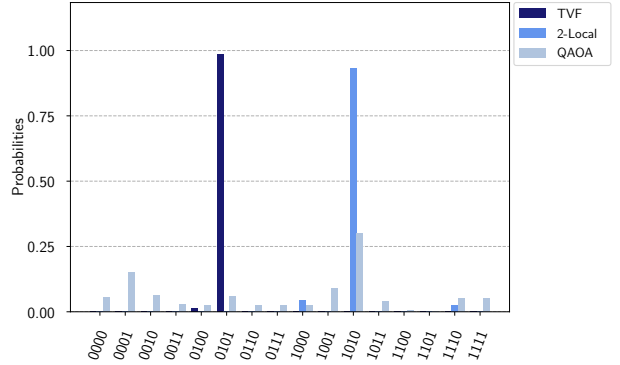}}
    \end{center}
    \caption{Comparison histogram solution for the FLP instance (eqs. \eqref{eq:facility_exampleobj} to \eqref{eq:facility_example3}), using TVF, 2-Local and QAOA.}
    \label{fig:hist_com_facility}
\end{figure}

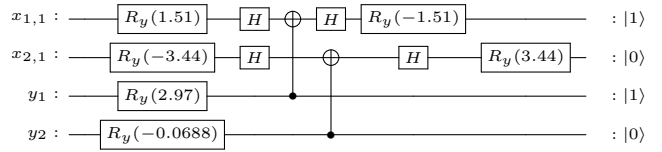
\begin{figure}[h]
\centering
\tiny
   \begin{equation*}
    \Qcircuit @C=0.8em @R=0.5em @!R {
	 	\lstick{ {x}_{1,1} :  }   & \qw & \gate{R_y(1.51)} & \gate{H}  & \targ & \gate{H} & \gate{R_y(-1.51)} & \qw& \qw & \rstick{ : \ket{1}  } \\
	 	\lstick{ {x}_{2,1} :  } & \qw & \gate{R_y(-3.44)}& \gate{H} & \qw & \targ & \gate{H} & \gate{R_y(3.44)} & \qw &\rstick{ : \ket{0}  }  \\
	 	\lstick{ {y}_{1} :  } & \qw & \gate{R_y(2.97)} & \qw & \ctrl{-2} & \qw & \qw & \qw & \qw & \rstick{ : \ket{1}  } \\
	 	\lstick{ {y}_{2} :  } & \qw & \gate{R_y(-0.0688)} & \qw & \qw & \ctrl{-2} & \qw & \qw & \qw & \rstick{ : \ket{0}  }\\
	 }
\end{equation*}
    \caption{Optimal circuit for the FLP instance \eqref{eq:facility_exampleobj} to \eqref{eq:facility_example3} using proposed methodology.}
    \label{fig:facility_circuit}
\end{figure}

In Figure~\ref{fig:hist_com_facility} we observe the most probable solution that correspond to the state $\ket{1010}$, this solution means that facility 1 is open ad delivers to the only one client.  At Figure~\ref{fig:facility_circuit} we observe the optimal circuit with set of final parameters $\mathbf{\theta} = \left(1.51,-3.44,2.97,-0.0688 \right)$.

\begin{figure}[h]
    \begin{center}
   \resizebox{\linewidth}{!}{\input{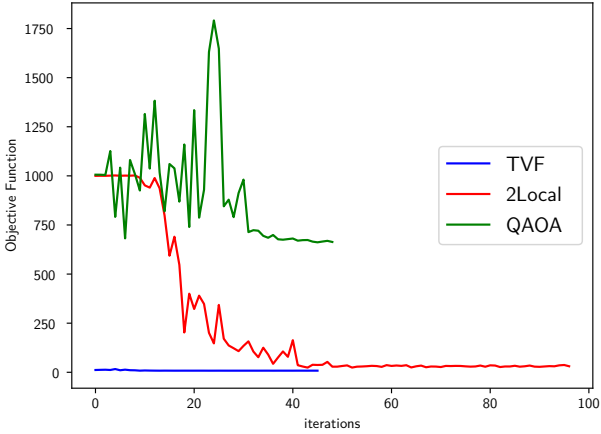}}
    \end{center}
    \caption{Comparison of evolution of the objective function for the FLP instance (eqs. \eqref{eq:facility_exampleobj} to \eqref{eq:facility_example3}), using TVF, 2-Local and QAOA.}
    \label{fig:iters_com_facility}
\end{figure}

COBYLA takes 45 iterations to get the final objective function 8.0. In Figure~\ref{fig:iters_com_facility} we observe the evolution of the Energy state for the different approaches. We observe that since we use less penalization functions for represent equality constraints while use the proposed methodology the Energy state starts in small values compared with the other approaches, facilitating the convergence to the objective function. Another important part of the validation of our methodology is the verification of the cost of the Ansatz implementation using all methodologies testes, as we observe the proposed has less cost compared with the others, we can warranty that for larger instances of the FLP problem since 2Local Ansatz increases the number of cnots with the number of qubits. In contrast, the number of parameters are certainly less than the other methodologies, facilitating the convergence of the classical optimization part.

\begin{table}[h]
    \centering
    \begin{scriptsize}
    \begin{tabular}{|l|c|c|c|c|}
\hline
  Var. Form   &  \# $SU(2)$  & \# cnots & \# param. & cost\\
  \hline
    TVF & 10 & 2 & 4 & 30 \\
    2-Local & 14 & 3 &  8 & 44 \\
    QAOA & 26 & 12 & 4 &  146\\
\hline
    \end{tabular}
    \end{scriptsize}
    \caption{Number of cnots and $SU(2)$ gates and parameters for tested circuits of TVF, 2-Local, and QAOA to solve FLP instance \eqref{eq:facility_exampleobj} to \eqref{eq:facility_example3}.}
    \label{tab:costs_FP}
\end{table}

\subsection{Linear Assignment Problem}

The Multi dimensional Assignment Problem (MAP) is a classical NP-Hard combinatorial optimization problem (\cite{10.1007/s10878-012-9554-z}), but for sake of simplicity and as a demonstrative example on how to use Tailored variational forms, we study the Linear Assignment Problem (LAP).

We consider a job assignment to a worker problem which is a LAP. Given a set of $n_{1}$ jobs and $n_{2}$ workers with more workers that jobs to be executed. Every job $i$ has a cost when a worker $j$ do it $c_{i,j}$. The purpose of this assignment is that every job will be done for a worker with a minimum total cost. Considering that every worker can do only one job.

The LAP can be formulated as presented in equations \eqref{eq:objLAP} to \eqref{eq:cons_LAP3}.

\begin{align}
    \max_{x} & & \sum_{i=1}^{n_{1}} \sum_{j=1}^{n_{2}} c_{i,j}x_{i,j} \label{eq:objLAP} \\
    s.t. & &\sum_{j=1}^{n_{2}} x_{i,j} = 1 & \; \; ~~ i = 1,\ldots,n_{1} \label{eq:cons_LAP1} \\
    & &\sum_{i=1}^{n_{1}} x_{i,j} \leq 1 & \; \; ~~ j = 1,\ldots,n_{2} \label{eq:cons_LAP2} \\
         & & x_{i,j} \in \{0,1 \} & \; \; ~~ i = 1,\ldots,n_{1} ; j = 1,\ldots,n_{2} \label{eq:cons_LAP3}
\end{align}

where variable $x_{i,j}$ is 1 if job $i$ is done by worker $j$ and 0 otherwise.

For LAP proble, all variables $x_{i,j}$ with the same $i$ index in constraint \eqref{eq:cons_LAP2} can directly use Figure~\ref{fig:circuit_disjunctive}, since $x_{i,j}$ variable only appears on one constraint, this is one limitation on this methodology. we use penalization for represent constraint \eqref{eq:cons_LAP1} as was used in FLP problem. Then is possible to formulate the algorithm~\ref{al:LAP} to construct the Ansatz for LAP problem. 

\begin{algorithm}[H]
    \caption{Linear assignment problem circuit construction}
    \label{al:LAP}
    \begin{algorithmic}[1]
    \REQUIRE number of jobs $n_{1}$ and workers $n_{2}$
    \REQUIRE Costs $c_{i,j}$ of a job $i$ done by worker $j$
    \STATE set all qubits with state 0.
   \FOR{$j = 1,\ldots, n_{2}$}
     \FOR{$i =  1, \ldots, n_{1} $}
    \STATE set qubit label $x_{i,j}$
     \STATE apply $R_{y}(\theta_{x_{i,j}})$ on  $x_{i,j}$
     \IF{i>1}
    \STATE apply $CX$ with $x_{i-1,j}$  as a control and $x_{i,j}$  as target
    \STATE $R_{y}(\theta_{x_{i,j}})$
    \ENDIF
    \ENDFOR
    \FOR{$i =  n_{1}, \ldots, 2 $}
    \STATE apply $CX$ with $x_{i-1,j}$  as a control and $x_{i,j}$  as target
    \ENDFOR
    \STATE apply $X$ on  $y_{i}$
    \ENDFOR
\end{algorithmic}
\end{algorithm}

To test the proposed methodology for the LAP, we solved the following instance: 

\begin{align} 
    \max_{x} & \;\; 5x_{1,1}+8x_{1,2} + 7x_{2,13}+11x_{2,2} \label{eq:assig_exampleobj} \\
    s.t. &  \;\; x_{1,1}+x_{1,2} = 1  \\
    &  \;\; x_{2,1}+x_{2,2} = 1  \\ 
    &  \;\; x_{1,1}+x_{2,1} \leq 1  \\
    &  \;\; x_{1,2}+x_{2,2} \leq 1  \label{eq:assig_example4}
\end{align}

We implement this instance of the FL problem with 4 qubits $(q_{1} \leftarrow x_{1,1}, q_{2} \leftarrow x_{1,2}, q_{3} \leftarrow x_{2,1}, q_{4} \leftarrow x_{2,2})$.  

\begin{figure}[h]
    \begin{center}
   \resizebox{\linewidth}{!}{\input{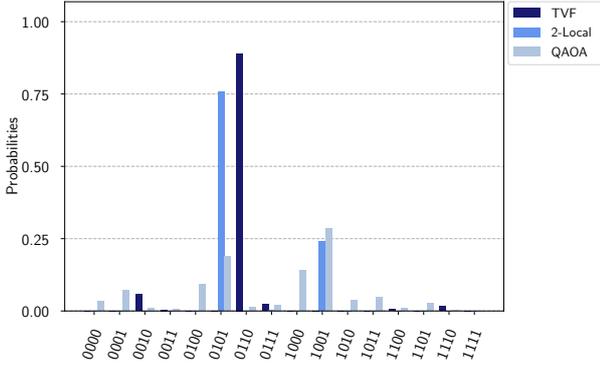}}
    \end{center}
    \caption{Comparison histogram solution for the LAP instance (eqs. \eqref{eq:assig_exampleobj} to \eqref{eq:assig_example4}) , using TVF, 2-Local and QAOA.}
    \label{fig:hist_assig}
\end{figure}

\begin{figure}[h]
\centering
\tiny
\begin{equation*}
    \Qcircuit @C=0.5em @R=0.5em @!R {
	 	\lstick{ {x}_{1,1} :  } & \gate{R_y(0.01)} & \ctrl{2} & \qw & \qw & \ctrl{2} & \qw & \qw & \qw\\
	 	\lstick{ {x}_{1,2} :  } & \gate{R_y(-3.22)} & \qw & \ctrl{2} & \qw & \qw & \ctrl{2} & \qw & \qw\\
	 	\lstick{ {x}_{2,1} :  } & \gate{R_y(-1.68)} & \targ & \qw & \gate{R_y(-1.68)} & \targ & \qw & \qw & \qw\\
	 	\lstick{ {x}_{2,2} :  } & \gate{R_y(3.42)} & \qw & \targ & \gate{R_y(3.42)} & \qw & \targ & \qw & \qw\\
	 }
\end{equation*}
    \caption{Optimal circuit for the set LAP instance (eqs. \eqref{eq:assig_exampleobj} to \eqref{eq:assig_example4}), using TVF.}
    \label{fig:assig_circuit}
\end{figure}
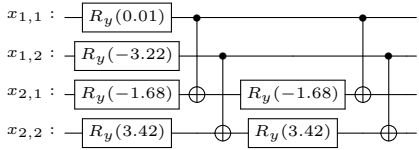

In Figure~\ref{fig:hist_assig} we observe the most probable solution that correspond to the state $\ket{0110}$, this solution means that job 1  will be done by worker 2 and job 2  will be done by worker 1.  At Figure~\ref{fig:assig_circuit} we observe the optimal circuit with set of final parameters $\mathbf{\theta} = \left( 0.01,-3.22,-1.68,3.42\right)$.

\begin{figure}[h]
    \begin{center}
   \resizebox{\linewidth}{!}{\input{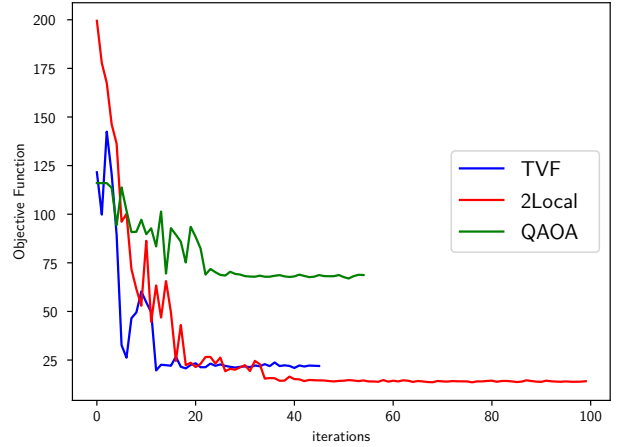}}
    \end{center}
    \caption{Comparison of evolution of the objective function for the LAP instance (eqs. \eqref{eq:assig_exampleobj} to \eqref{eq:assig_example4}), using TVF, 2-Local and QAOA.}
    \label{fig:iters_assig}
\end{figure}

For the LAP instance that we investigate, COBYLA takes 45 iterations to get the final objective function 15 . As we see in Figure~\ref{fig:iters_assig}, TVF takes less iterations to converge to the optimal solution, while the other converge to infeasible solutions as is depicted in Figure~\ref{fig:hist_assig}. In this case computational cost is slightly more than 2Local Ansatz as ins presented in Table~\ref{tab:costs_LAP2}.

\begin{table}[h]
    \centering
    \begin{scriptsize}
    \begin{tabular}{|l|c|c|c|c|}
\hline
  Var. Form   &  \# $SU(2)$  & \# cnots & \# param. & cost\\
  \hline
    TVF & 6 & 4 & 4 & 46 \\
    2-Local & 14 & 3 &  8 & 44 \\
    QAOA & 24 & 8 & 4 &  104\\
\hline
    \end{tabular}
    \end{scriptsize}
    \caption{Number of cnots and $SU(2)$ gates and parameters for tested circuits of TVF, 2-Local, and QAOA to solve LAP instance (eqs. \eqref{eq:assig_exampleobj} to \eqref{eq:assig_example4}).}
    \label{tab:costs_LAP2}
\end{table}

\section{Conclusion} \label{sec:Conclusions}
One of the main advantages of our proposed tailored variational forms is that with this circuits we do not explore the entire Hilbert space, we only explore the points where the solutions are feasible, since explore the entire Hilbert space could imply in more computational cost from the classical optimization part \cite{McClean_2018}. As was observed the construction of a TVF has a constant number of parameters for a given number of qubits, this methodology have no meta-parameters to decide, those parameters usually multiply the number of parameters of a block of entanglers and fixed the number of rotation gates. One limitation is that we do not have a general receipe to combine this TVFs in order to represent multiple constraints on a LCQBO problem. As a future work is possible approximate the TVFs by reducing in circuits on blocks with a predefined depth, in order to use less cnots gates. Another possible development is that with expressions \eqref{eq:validstates1} and \eqref{eq:validstates2} we can calculate the gradient with respect to the parameters $\theta$, and use gradient based methods, we didn't explore this possibility, since for our small problems, COBYLA solver converges quickly.

\section{Acknowledgments}
We would like to praise Takashi Imamichi (IBM Research) for his insightful comments and suggestions on this manuscript.

\bibliographystyle{plainnat}
\bibliography{main.bbl}

\end{document}